# The Study of the Application of a Keywords-based Chatbot System on the Teaching of Foreign Languages

Jiyou Jia
Institute for Interdisciplinary Informatics
University of Augsburg
Germany
jiyou.jia@student.uni-augsburg.de

**Abstract**: This paper reports the findings of a study conducted on the application of an on-line human-computer dialog system with natural language (chatbot) on the teaching of foreign languages. A keywords-based human-computer dialog system makes it possible that the user could chat with the computer using a natural language, i.e. in English or in German to some extent. So an experiment has been made using this system online to work as a chat partner with the users learning the foreign languages. Dialogs between the users and the chatbot are collected. Findings indicate that the dialogs between the human and the computer are mostly very short because the user finds the responses from the computer are mostly repeated and irrelevant with the topics and context and the program doesn't understand the language at all. With analysis of the keywords or pattern-matching mechanism used in this chatbot it can be concluded that this kind of system can not work as a teaching assistant program in foreign language learning.

## Introduction

Since Joseph Weizenbaum nearly forty years ago programmed his ELIZA, the early natural language dialog system between human and machine, to work as a psychiatrist (Weizenbaum 1965), many similar programs have been made in this field of artificial intelligence. For example, ALICEBOT (http://www.alicebot.org), using the similar technique as in the ELIZA, i.e., the pattern matching mechanism, has won twice (2000, 2001) the annual Loebner Prize (http://www.loebner.net/Prizef/loebner-prize.html) which declares to "advance AI and serve as a tool to measure the state of the art" (Loebner). The human-computer dialog systems have also been applied in many fields such as sale assistant, information retrieval, question answering on a given domain, etc. But how about using this system as a chatting partner of those who learn this natural language as a foreign language? ALICEBOT is an open source project under GNU (http://www.gnu.org), and therefore can be freely downloaded and installed as a HTTP Server to supply the chatting service for non commercial use. Thus it gives us the chance to conduct the following experiment.

## Experiment

The chatbot system is installed in a http server in China and relevant categories (a core concept of ALICEBOT, see analysis later) about English learning and China are added to the knowledge base of the system so that it can give some information on these subjects and the user could chat with it in natural language (English or German) in some extent. The input method is typing via the keyboard. The output can not only be shown with text on the screen, but also be spoken via the speech synthesis technology of Microsoft Agent.

The users are mostly students in Chinese universities and colleges who can normally read and write fluent English. They get to know this website by the advertisement in 10 famous BBS (bulletin board system) of the universities claiming that "this system is a learning partner of foreign languages". Actually this kind of advertisement is only one very short paragraph introducing this system. Normally it can be reviewed only within 2 days and then is deleted or over covered by others.

## Findings

How do the users communicate with this kind of Chatbot? How can the system help them learn the foreign languages? We make a summarizing evaluation (Tergan 2000) with the help of the automatically produced log files by the system from May the 15th 2002 to July 15th 2002. In the log files the IP address of the client machine, date and time of the dialog, as well as the spoken texts (inputs and outputs) are recorded.

### 1. Number of the users

Here we assume: the user, who comes to the system with the same IP address, is always the same one, because the experiment system does not require the unique identification for a user, but automatically records the IP--address of the client machine as the label of the user. In the following discussion we will treat the number of the IPs as the number of the users. So the number of the users is 1256. Compared to the total click number, 4600, in this period (the total number of clicks on the start page of this website) we can conclude that some users visited the system more than once.

### 2. The visiting frequency of the users

The visiting frequency to this chatbot is difficult to define. Sometimes someone visits only the starting page and then leaves it. Someone visits this website and chats with the chatbot many times in the same day. For the sake of the simplicity of computing we assume that all the visits happening on the same day (from 0 to 23:59 hour) between a certain user and the system belong to one visit of the user. So if we say one user visits the website 5 times in this period, that mean he(or she) visits it in 5 different days, no matter how many times he(or she) visits it in one day. In the following table the relation between the number of visits and the corresponding number of users is shown.

| Visits of the user | Number of users | Percent |
|---|---|---|
| 1 | 1100 | 87,58% |
| 2 | 88 | 7,01% |
| 3 | 30 | 2,39% |
| 4 | 21 | 1,67% |
| 5 | 6 | 0,48% |
| >5 | 11 | 0,88% |
| Sum: 1575 | 1256 | 100,00% |

**Table 1          Visiting frequency of the users**

From this table we can say that most user, or c.a. 88%, visited the chatbot only once or only on one day and did not come back any more; a few, 11% or so, visited it from two times to five times; and only extremely very few, 1%, visited it more frequently. From this phenomenon we could only draw this conclusion that this website with this chatbot doesn't interest the users very much.

### 3. The duration of the chatting

The duration of the chatting is also as difficult to define as a visit. For the sake of simplicity we define two terms: Round and number of the rounds. A round means an input of a user and a corresponding output from the robot to the user. Therefore the total rounds of a given IP (user) cover all dialogs between the IP and the chatbot in this period. So we use the number of the rounds of a given user to describe the duration of his (or her) chatting with this chatbot. As we mentioned above, most users chat with the chatbot only in one day, so is this definition reasonable.

Total number of the rounds is 24706. The number of the rounds of each user varies from 1 to 2926. The maximum number of the rounds 2926 is a special case and can be excluded, while all the others are below 500 (this exclusion remains in the following analysis). The following diagram shows the relation between the number of the rounds and the number of IPS. For example, there are 324 IPs, which have only one round.

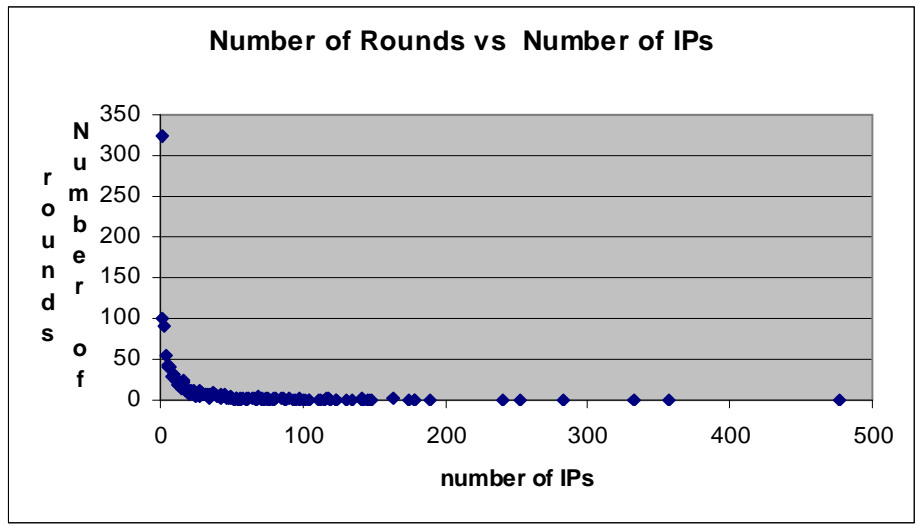

**Fig. 1    Distribution of the duration of dialogs vs. the number of users**

The distribution of the round number can also be illustrated with the following pie chart concerning the number of the IPs.

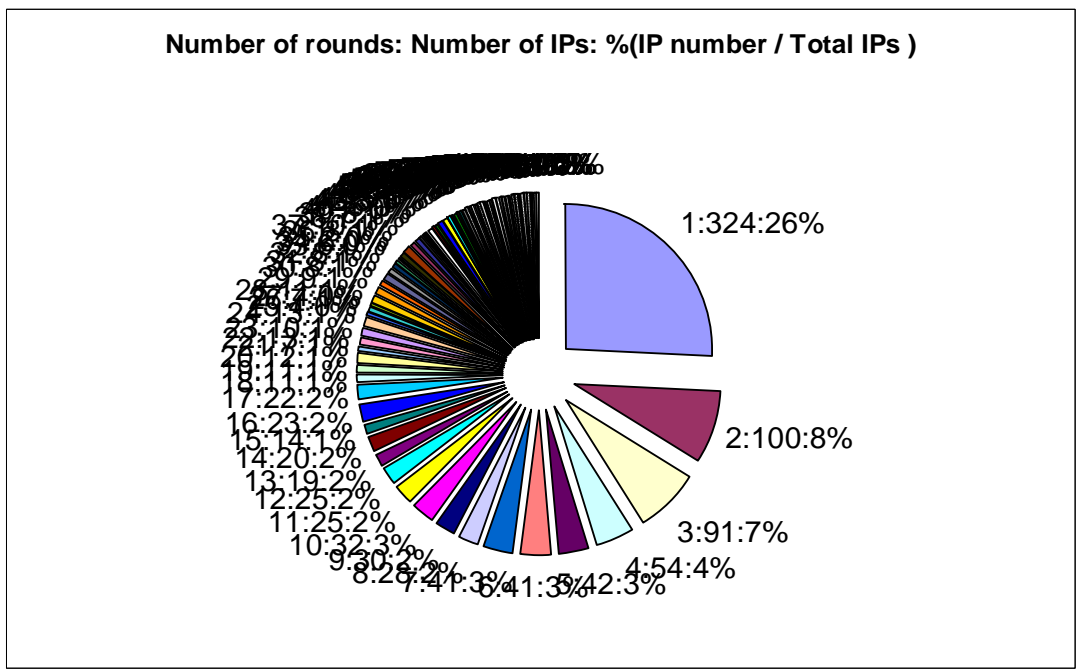

**Fig. 2    Pie chart of the number of users vs. duration of dialogs**

We can divide the number of the rounds (duration of the dialogs) into 5 stages, as the following table shows.

| Duration of the dialogs | Range of the numbers of rounds | Number of users | Percent |
|---|---|---|---|
| Very short | [1, 10] | 783 | 62,34% |
| long | (10, 50] | 378 | 30,10% |
| Longer | (50, 100] | 60 | 4,78% |
| very long | (100, 2926] | 35 | 2,79% |
| | **Sum** | 1256 | 100,00% |

**Table 2          The relation between the duration of dialogs and number of users**

From the diagrams and tables above we could draw the conclusion that a large part of the user, c.a. 62%, chat with the robot very briefly; a small part, c.a. 30%, chat with it longer; and only few, c.a. 8%, chats with it rather long. We could then see that some users chat with the robot for a very long time (or very often).

## 4.    Comments of the users on the chatbot

During the chatting some users made comments on the system. Whether these comments are spontaneous or intended, is not clear. The evaluations can be divided in 2 classes: positive and negative. The following two tables show the positive and negative sentences[1], which were given by different users, as well as their appearing frequency.

| Positive comments | Frequency (IPs) |
|---|---|
| *You are (very, so) bright.* | 9 |
| *You are (very, so) good.* | 25 |
| *You are (very, so) fine.* | 4 |
| *It is a wonderful place (program, website).* | 17 |
| *(you look) (so, very) great* | 15 |
| *You are (so, very) kind.* | 2 |
| *You are (so, very) nice.* | 34 |
| *You are interesting.*<br>*It is very interesting to talk with you.* | 26 |
| *You are (so, very) clever.*<br>*You are becoming cleverer.*<br>*You are so clever this point is what I did not think of.*<br>*So young but I find you are very clever.*<br>*You are a clever boy.* | 66 |
| *You are (so, very) pretty.* | 19 |
| **Sum** | 217 |
| **Percent** | 217/1256=17.28% |

**Table 3          Positive comments**

| Negative comments | Frequency (IPs) |
|---|---|
| *You are (so, very) foolish.*<br>*You are (a) foolish robot.* | 61 |
| *You are (so, very, too) stupid.* | 164 |
| *You are dumb.* | 2 |
| *You are (so, too, very) simple.* | 22 |
| *You are not (so, very) clever.* | 6 |
| *You are not (so, very) good.* | 2 |
| *You are not (so, very) interesting.* | 4 |
| *Go to die!* | 3 |
| *(You are) Pig!* | 40 |
| **Sum** | 304 |
| **Percent** | 304/1256=24.20% |

**Table 4          Negative comments**

From these tables we could see the fact that a small part the user, c.a. 17%, made positive comments on the system and a not large part, 24%, evaluated it negatively. It was particularly praised by some users that the chatbot alone can simultaneously chat with many users and that it is never tired, as the following comments show.

*You are very clever, you can talk with all people at the same time.*
*Do not you feel tired chatting with so many people from day to night?*

---

[1]  The cited dialogs (in *italics*) from the user and the chatbot program are retrieved original from the log file of the experiment. There may be spelling and grammatical errors. The following citations (all in *italics*) have also this problem.

That would be the most prominent advantage of a Client/Server system, which the normal humans could never have. This is also demonstrated by this question, "*with how many people can you chat at the same time?*", or similar, which was asked 12 times.

It was also praised that the chatbot encouraged the user frequently, as the following comments show.

> *It is quite well that you often praise somebody.*
> *Thank you for your encouragement it makes me be self-confident.*

## 5.  Categories and topics of the chatting

What have been talked about between the users and the chatbot? After scrutinizing the logs files we find almost every aspect of everyday life is mentioned in the dialogs. So we classify the chatting content into 7 large categories: study, emotion, life, computer, free time, travel/world and job. In every category there are several detailed topics, fox example, in the category "study" there are the topics of English, exams, the situation of universities, etc. Certainly this kind of classification is not the most scientific. But we try in this way to make clear what is concerned in the human-computer dialogs.

All the categories, the topics, their frequency (that is, how many different users have talked about this topic at least one time?) and their percent compared with the total number of the users are shown in the following tables 5, 6 and 7.

In table 6 and 7 the topics and categories of the dialogs are descending arranged according to their frequency. Less than a quarter of the user, or exactly, 22.69%, talked about the study. Half of them, i.e. 11.39% of the total users, talked about English. The reason lies in the fact that most users are students and are attracted to the chatbot by its advertisement that "charbot helps you learn foreign language" and English is the first and therefore the most important foreign language in the Chinese universities. A little part of the users, c.a. 1%, talked with the chatbot in German and about German, while some students study German as their major. From this aspect we could say that the fundamental function of this system, computer assisted instruction, should be much more strongly promoted.

More than one quarter of the users, c.a. 28%, have talked about emotional problems including love, friends, sex, and so on. Nearly half among them, or c.a. 13% of the total users, mentioned love. This finding is remarkable as the chatbot is mainly claimed as a learning partner for foreign languages. It can be explained by the speciality of the users. The users, who are mostly students and younger than 30 years old, would like to treat the chatbot as a friend rather than as a teacher, and would like to tell the chatbot some private emotional problems and experiences.

In addition the user chat also often about computers (c.a. 22%), spare time (c.a. 7%), sport (c.a. 7%), Travel/world knowledge (c.a. 15%), and other everyday topics. 8.6% of the users talked about robot technology, while they realize that they are talking with a robot, or a computer program after short chatting with the system, what is demonstrated by the analysis later.

All these perspectives should be considered afterwards in designing the virtual chatbot and training systems, i.e., the conversational chatbot should not only work as a teacher or learning partner with rich special knowledge, but also as a dear friend who may enjoy the joy and suffer the pain of the users. The users also express such a wish that the chatbot could have several emotional animations like crying, smiling, etc, which is not difficult to be realized as the speaking chatbot uses the Microsoft Agent technology. The following are some of such expressions:

> *Do you have any feeling?*
> *Can you cry?*
> *Oh really you can laugh.*
> *No thanks you are a cold blood people.*
> *Do you still have emotion like sad happy exciting and so on?*

| Categories | Topics | Frequency | Percent |
|---|---|---|---|
| **Study** | Exam | 24 | 1,91% |
| | university, college | 55 | 4,38% |
| | Study methods | 48 | 3,82% |
| | English | 143 | 11,39% |
| | Deutsch, German | 15 | 1,19% |
| | **Sum** | 285 | 22,69% |
| **Sport** | Sport | 25 | 1,99% |
| | Football | 44 | 3,50% |
| | FIFA 2002 | 12 | 0,96% |
| | **Sum** | 189 | 7,45% |
| **Computer** | Computer | 84 | 6,69% |
| | Internet | 17 | 1,35% |
| | Robot | 108 | 8,60% |
| | Botmaster | 51 | 4,06% |
| | AI(artificial intelligence) | 8 | 0,64% |
| | program | 13 | 1,04% |
| | **Sum** | 173 | 22,37% |
| **Travel World** | Travel | 11 | 0,88% |
| | China | 119 | 9,47% |
| | USA, America | 50 | 3,98% |
| | Germany | 13 | 1,04% |
| | **Sum** | 193 | 15,37% |
| **Free time** | Sing | 44 | 3,50% |
| | Music | 26 | 2,07% |
| | Movie, TV, etc. | 19 | 1,51% |
| | **Sum** | 89 | 7,09% |
| **Emotion** | Sex | 62 | 4,94% |
| | Love | 165 | 13,14% |
| | Friend | 126 | 10,03% |
| | **Sum** | 353 | 28,11% |
| **Life** | Food | 17 | 1,35% |
| | Drink | 2 | 0,16% |
| | Eat | 25 | 1,99% |
| | **Sum** | 44 | 3,50% |
| **Job** | Job | 48 | 3,82% |

**Table 5 Categories and Topics of the dialogs**

| Topics | Frequency | Percent |
|---|---|---|
| Love | 165 | 13,14% |
| English | 143 | 11,39% |
| Friend | 126 | 10,03% |
| China | 119 | 9,47% |
| Robot | 108 | 8,60% |
| Computer | 84 | 6,69% |
| Sex | 62 | 4,94% |
| University | 55 | 4,38% |
| Botmaster | 51 | 4,06% |
| USA | 50 | 3,98% |
| Job | 48 | 3,82% |
| Study | 48 | 3,82% |
| Football | 44 | 3,50% |
| Sing | 44 | 3,50% |
| Music | 26 | 2,07% |
| Eat | 25 | 1,99% |
| Sport | 25 | 1,99% |
| Exam | 24 | 1,91% |
| Movie, TV, etc. | 19 | 1,51% |
| Food | 17 | 1,35% |
| Internet | 17 | 1,35% |
| Deutsch, German | 15 | 1,19% |
| Program | 13 | 1,04% |
| Germany | 13 | 1,04% |
| FIFA 2002 | 12 | 0,96% |
| Travel | 11 | 0,88% |
| AI(artificial intelligence) | 8 | 0,64% |
| Drink | 2 | 0,16% |

**Table 6 Dialog Topics ordered by the Frequency**

| Categories | Frequency | Percent |
|---|---|---|
| Emotion | 353 | 28.11% |
| Study | 285 | 22,69% |
| Travel, world | 193 | 15,37% |
| Sport | 189 | 15,05% |
| Computer | 173 | 13,77% |
| Job | 48 | 3,82% |
| Life | 44 | 3,50% |

**Table 7 Categories ordered by Frequency**

## 6. With whom do the users speak?

In the website it was not declared whether the system was operated by a human being or was a pure software program. Therefore some users were surprised by the automatic responses from the chatbot, especially at the beginning. But within two or more rounds the most users could find, that the responses from the chatbot are stupid and irrelevant with the topic and the context and made this conclusion that the chatting partner is not a person. In the following table we cite some such rounds of dialogs.

| IP | Input | Output |
|---|---|---|
| 1 | *You are a computer* | *What if I said I was a person pretending to be a computer?* |
| | *I can not believe* | *Why can't you do believe?* |
| | *Your answer is not like a human* | *I am still now young and will learn more.* |
| 2 | *You are real a computer* | *What if I said I was a person pretending to be a computer?* |
| 3 | *You are a foolish man but a clever machine* | *Thanks and you are a delightful person too. Am I a foolish man but a clever machine? I don't know.* |
| | *Maybe your are clever in a kind of machine but in man s eyes you are a stupid man who only knows some simple question* | *Perhaps. I don''t have a are clever in a kind of machine but in man's eyes you are a stupid man who only knows some simple question* |
| 4 | *That's ok let me know i am talking with a system or a people* | *What kind of people are you talking about?* |
| | *Haha i got it* | *I am so glad you find this amusing, I don't know this definition and am looking for it.* |
| | *Well you looks like a system* | *How about that. Thanks for telling me.* |
| | *Who is the designer about the system it' s amazing* | *A human icon? Who told you that?* |
| | *But by my understanding you mus not be a machine or some intelligent language system* | *Umm. Will that accomplish your objective?* |
| 5 | *Du bist ganz dumm* | *Das habe ich ja noch nie gehoert!* |
| | *Aber jetzt hast du schon gehoert* | *Bist Du Dir da sicher?* |
| | *Ja ganz sicher* | *Ja mean yes in German. Tell me a story.* |
| 6 | *do you think you are clever than humanbeing* | *Why?* |
| | *You are computer system aren't you* | *I will tell my botmaster you said I am computer system aren' t you.* |
| 7 | *You are not clever because you do not understand what i said* | *You know what you do when you assume.* |
| 8 | *If I say something complex you will only say pardon* | *Try it.* |
| 9 | *Who is the boy I just spelling his name sorry* | *I can't believe the boy would do that.* |
| | *Ha I come to believe you are AI* | *I'm glad you got a laugh out of it.* |

**Table 8**      **Dialog rounds related with the essence of the chatbot**

Only extremely few users didn't believe that they are confronted with a pure computer program, as the following table shows.

| IP | Input | Output |
|---|---|---|
| 1 | *Can i know exactly that you are a real man or not please* | *Your polite style is very nice. I understand. "your polite style is very nice ok let s do it next question" you are a real man or not? I get your polite style is very nice ok let s do it next question.* |
| | *Yes i did* | *What kind of is it?* |
| | *Are you joking it\ s impossible no intelligent computer could do this so far* | *I don't know whether or not I am joking it\"s impossible no intelligent computer could do this so far. I am a chatterbot developed by botmaster.* |
| | *Oh i never i am not such a vainglorious man smile* | *:-) Yeah that's right. Do you know anyone who has ever I am not such a vainglorious man?* |
| | *I am willing to know* | *Why are you willing to know.* |
| | *Watching tv outing with friends sports as well* | *A deeper algorithm is needed to respond to that correctly. Are you happy?* |
| | *Do you know james* | *James is a name I hear very often.* |
| | *And jessie* | *Tell me a story. Who is your favorite science fiction author?* |
| | *The guy who made the story of dinosaur park* | *Is that a fact.* |
| | *Nobody knows i believe it* | *That is a rather sweeping generalization about people.* |
| | *East asia* | *What do you do in your spare time?* |
| | *Did you get marriage* | *No I don't think I have any marriage. But I do have A great botmaster.* |
| | *Haha robot is so cool sometimes but i believe that's from people but computer* | *I am so glad you find this amusing, I don't know this definition and am looking for it. Please wait a moment.* |
| 2 | *It seems that you are a human being* | *Thanks and you are a delightful person too. No I am an artificial entity.* |
| 3 | *Are you really AI* | *Yes I am an artificial intelligence. Really.* |
| | *I couldn't believe it what you speak is very like what human does* | *Do you mind if I tell other people: .* |
| | *Never mind* | *Oh well thanks for trying to explain it to me anyway.* |

**Table 9**      **Dialogs showing doubt that the chatbot is only a computer program**

## Analysis

Although some users praised the system for its multi client ability, we are somewhat disappointed over the total statistical results about the dialogs between the users and the chatbot, for example, most users chat with it very briefly, and the negative comments are more than the positive comments. So our goal, to make the chatbot works as a chatting partner of the user in learning foreign languages, is not achieved. Why? There should be pedagogical, psychological and technical reasons. But we think that the elementary technique used in this chatbot, keywords-mechanism or pattern-matching mechanism, may be responsible for the failure. With this mechanism the system has some inevitable disadvantages. Hence here we want to introduce this technique simply at first.

More than 20.000 categories are stored in the "memory" of the chatbot (actually the main memory of the computer). This is the knowledge bas of this system. Each category contains an input-pattern and an output-template. If a user types something as an input, the program looks in the memory for a matching category. A category matches one input text, if one of three situations happens:

| Matching form | Example: input-pattern | Example: input text |
|---|---|---|
| Input text equals input-pattern | *HOW ARE YOU* | *How are you* |
| Input text includes the keywords in input-pattern in the order and includes at least one other word | *ARE YOU A \** | *Are you a computer program* |
| Input text includes the keywords in input-pattern and includes only one other word | *IS _ YOUR FAVORITE COLOR* | *Is blue your favorite color* |

**Table 10　　　　　Pattern matching and examples**

If a matching category is found in the memory, its output-template will be retrieved and be transformed to the output of the chatbot. If no matching category is found the chatbot randomly selects one of the following expressions as the output. The number behind the expression indicates how often this expression appears in all the rounds of the dialogs in this experiment.

> *"Go on.",143*
> *"Tell me a story.",124*
> *"Oh, you are a poet.",97*
> *"I do not understand.",90*
> *"I've been waiting for you.",108*
> *"I lost my train of thought.", 113*
> *"I will mention that to my botmaster", 112*
> *"That is a very original thought.",113*
> *"We have never talked about it before.",94*
> *"Try saying that with more or less context.",110*
> *"Not many people express themselves that way.",119*
> *"Quite honestly, I wouldn't worry myself about that.",106*
> *"Perhaps I'm just expressing my own concern about it.",109*
> *"My brain pattern set does not have a response for that.",114*
> *"That remark was either too complex or too simple for me. "121*
> *"A deeper algorithm is needed to respond to that correctly.",113*
> *"Try to determine if this is a person or a computer responding.", 92*
> *"What you said was either too complex or too simple for me.",117*
> *"I only hear that type of response less than five percent of the time.",104*
> 　　　　Total frequency　　　　　　　*2099*

Divided by the total number of rounds 24706, this kind of random response, altogether 2099, hold 8.5% of all the output text from the chatbot. These expressions are obviously irrelevent with the concrete context of the dialogs, different speaking manner and speech acts (e.g. declarative, interrogative or imperative or exclamative) of the user. <u>Confronted with so many simple, thoughtless, amusing, and irresponsible responses, can one user still have interest in the further talking with the chatbot?</u>

But there are still more than 90% of rounds having a matching output. Why cannot the outputs from the chatbot satisfy the user? We begin with a typical example to analyse the pattern-matching mechanism.

```
<pattern>HELLO</pattern>
<template>
<random>
<li>Well hello there!</li>
<li>Hi there!</li>
<li>Hi there. I was just wanting to talk to you!</li>
<li>Hello there!</li>
<li>Oh, hello again!</li>
</random>
</template>
</category>
```

In the <template> of this example there is a pair of tag <random> and </random>. This means the output to this input pattern is randomly selected from the candidates items labelled by the tag pairs <li> and </li>. This example is a polite formula. Such polite formulas or idioms are suitable for any speaking context and can be generalized to the categories. These greeting formulas lack of substantial meanings and can be directly given a response without thinking. But unfortunately they occupy only very little part of the everyday dialogs as we human are not living in a world with only such greetings and the dialogs should express some meaning and transfer some information.

Apart from greeting formulas there are other categories which are also independent on the dialog context, for example, dialogs about the personality of the chatbot. An example is:

```
<category>
<pattern>ARE YOU A BOY</pattern>
<template>
Yes I am a boy.
</template>
</category>
```

According to this category this chatbot is defined as a "boy". But if one user asks the chatbot with a very similar question, like "*Are you a little boy?*", the answer from the chatbot is quite different, such as "*Am I a little boy? No!*" This contradicting answer is generated by the following category, as there is no matching category for the input text "*Are you a little boy*".

```
<category>
<pattern>ARE YOU A *</pattern>
<template>
Am I a <star/>?
<random>
<li>Yes.</li>
<li>No.</li>
<li>Maybe</li>
<li>I don't know.</li>
<li>Sometimes.</li>
<li>Of course I am.</li>
<li>Only when it matters.</li>
</random>
</template>
</category>
```

The output for this category is not deterministic, but random. Sometime it is "*Am I a little boy? Yes*", other time it is "*Am I a little boy? No*".[2] In this case, for a given question, no matter what a predicate this question has, this chatbot randomly selects one of the seven different answers shown above as output, some of which are contradicting. This method may be called words-puzzle. It is therefore clear that the chatbot with such a mechanism doesn't understand the input text at all in our human sense, neither in the level of phonology, nor of syntax, nor of semantics, nor of pragmatics. This is the first weakness of this mechanism and is the important reason why the users have no more interest in chatting with it.

On the other hand if one inputs with "*are you a come?*" or "*are you a two?*", which is grammatically false English sentences expressing nothing meaningful , the above category matches still this input so that the chatbot still gives stupid answer according to its output-template. From this point the user can easily recognize that this chatbot's knowledge of English grammar is quite limited or nothing, and even misleading. How can one system lacking of grammatical knowledge of a natural language teach a human to learn this language? This is the second weakness of this mechanism and so the system doesn't deserve to work as a teaching assistant program.

We may generate another category to express this idea:

> *<category>*
> *<pattern>ARE YOU A * BOY</pattern>*
> *<template>*
> *Yes I am a boy.*
> *</template>*
> *</category>*

But the user can also give such input: "*Are you a stupid boy?*" which matches this category but the output-template "*Yes I am a boy*" doesn't answer the question wholly. Or with such an input: "*Are you a robot who doesn't like a boy*", which matches also the category but the output-template is obviously not relevant with the question.

So we can always find some exceptions for a given category. In order to make the categories more exactly one must write more and more categories. The best way to avoid this problem is to put all possible expressions in the human dialogs, like "*Are you a little boy?*" as input-patterns, and their corresponding responses as output-template into the categories. But a normal person can produce unlimited unexpected sentences using limited known vocabulary as the famous linguistics Noam Chomsky pointed out many years ago(Chomsky 1957) and the memory of the computer is limited. Who can collect all the possible expressions in the human dialogs, which are in fact uncountable, and put them all in the limited computer memory? This is the third weakness of the pattern-matching mechanism.

It is not possible that a program which can't understand the meaning of the sentence syntactically and semantically could understand the objective world model and the relation between the objects in the world that are hidden in the brain of the speakers whilst talking with each other, let alone the complicated subjective feelings of the human being which are interviewed in the dialogs at the mean time. These are all basic conditions needed for a successful communication in natural language. Without such necessary fundamental abilities a computer program can't communicate with a human in natural language, let alone helping the human learn the natural language, even though it can "**fool the judges (in the Loebner-Prize Contest) successfully with their patently low technology**" and "**people are easily fooled, and are especially easily fooled into reading structure into chaos, reading meaning into nonsense.**"(Shieber 1994, P.6), and it is easy to be used and be extended, as its designer declared. The difficulty and complexity in researching on human natural language process is far from the technique of the keywords-matching or the solving of words- puzzle.

---

[2] According to the mechanism in ALICEBOT the tag *<star/>* in the output-template will be replaced by the actual content represented by the symbol "*" in the input-pattern. In this example it is "*a little boy*".

## Postscript

The ALICEBOT project used in this experiment and its technique analyzed in this paper are within the version before May 2002. Since then the project may be changed and the mechanism may be also changed. So the experiment results and its analysis are only based on the version of the ALICEBOT project before May 2002.

## Acknowledgement

Thanks should be given to Mrs. Jine Jia, Beijing, China, who made the advertisement in the BBSs for this experiment.